\documentclass[conference,onecolumn,12pt]{IEEEtran}

\usepackage{enumerate}
\usepackage{algorithm,amsmath}
\usepackage{algpseudocode}
\usepackage{amssymb}
\usepackage{color}
\usepackage{graphicx}
\usepackage{hyperref}
\usepackage{amsmath}
\usepackage{subfigure}
\usepackage{multirow}

\newtheorem{Lemma}{Lemma}

\usepackage{lipsum}
\newcommand\blfootnote[1]{%
  \begingroup
  \renewcommand\thefootnote{}\footnote{#1}%
  \addtocounter{footnote}{-1}%
  \endgroup
}

\begin{document}

\title{Multi-UAV Data Collection Framework for Wireless Sensor Networks}

\author{\IEEEauthorblockN{Safwan Alfattani, Wael Jaafar, Halim Yanikomeroglu, Abbas Yongacoglu}
}


\maketitle

\begin{abstract}
In this paper, we propose a framework design for wireless sensor networks based on multiple unmanned aerial vehicles (UAVs). Specifically, we aim to minimize deployment and operational costs, with respect to budget and power constraints. To this end, we first optimize the number and locations of cluster heads (CHs) guaranteeing data collection from all sensors. Then, to minimize the data collection flight time, we optimize the number and trajectories of UAVs. Accordingly, we distinguish two trajectory approaches: 1) where a UAV hovers exactly above the visited CH; and 2) where a UAV hovers within a range of the CH. The results of this include guidelines for data collection design. The characteristics of sensor nodes' K-means clustering are then discussed. Next, we illustrate the performance of optimal and heuristic solutions for trajectory planning. The genetic algorithm is shown to be near-optimal with only $3.5\%$ degradation. The impacts of the trajectory approach, environment, and UAVs' altitude are investigated. Finally, fairness of UAVs trajectories is discussed.
\end{abstract}


\blfootnote{We would like to thank Arslan Ahmed and Dr. Faraj Lagum for the valuable discussions. This work is supported in part by the Natural Sciences and Engineering Research Council Canada (NSERC), and in part by a scholarship from King AbdulAziz University, Saudi Arabia.}
\vspace{-10pt}
\section{Introduction}
In wireless sensor networks (WSNs), a large number of sensor nodes (SNs) is usually deployed to detect physical phenomena, e.g., pressure, temperature, etc. 
As part of the Internet of Things (IoT), WSNs are everywhere today, enabling new applications including smart grids, water networks, and intelligent transportation. In these systems, SNs are low-power devices with typically short transmission ranges. Current IoT SNs are based on IEEE 802.15/802.11af and have a maximum communication range of few hundred meters. In such systems, data sensed by SNs needs to be transmitted to a network manager. Typical WSNs deploy sinks (gateways) to collect SNs data and forward it to the Internet. However, due to the limited range of SNs, large numbers of sinks are deployed, which raise capital and operating expenses. 

Recently, low altitude platforms based on unmanned aerial vehicles (UAVs) have been presented as a promising technology to
enable several wireless applications such as coverage extension, disaster/emergency situation management, etc. \cite{Irem2016,Alzenad2017}. In the context of WSNs, UAVs can act as data collectors from SNs. In \cite{Pang2014}, with SNs partitioned into clusters, UAVs were used to collect data from them. However, this work was limited to static networks and one type of environment. Also, optimal deployment of UAVs has not been investigated. Authors in \cite{Mozaf2016,Zeng2018_1} investigated the deployment and trajectory of a single UAV supporting downlink wireless communications only. These works are limited since they focused only on one UAV and downlink. In \cite{Zeng2019}, a single UAV was used to communicate with ground nodes. The UAV's trajectory was optimized to minimize power consumption. However, multiple UAVs were not considered. \cite{Mozaf2016_1} proposed clustering to find optimal locations and trajectories of UAV cells to maximize collected data from SNs. Nevertheless, in large WSNs, UAVs may be constrained by hovering for long periods of time above each cluster to collect all data, or some SNs data may be dropped due to UAV power limitations.      

Most studies investigated single UAV deployments and ignored multi-UAV cases. Moreover, communication with low-power SNs is impractical, and hence deploying cluster heads (CHs) with higher power capabilities, in conjunction with UAVs, is recommended \cite{Afsar2014}. Also, most state-of-the-art research considered perfect Line-of-Sight (LoS) channels. This is not true in urban environments, where Non-LoS (NLoS) links exist. Being conscious of these limitations, we address here the problem of designing multi-UAV WSNs to enable data collection while minimizing the total flight time.

Our main contribution is as follows: we propose a WSN framework where CHs and their locations, and UAVs and their trajectories are optimized to collect WSN data in the shortest possible time. First, using K-means clustering, the number and locations of CHs are optimized, such that each SN can communicate reliably with its associated CH. Then, by leveraging optimal and heuristic solutions of the multiple Travelling Salesman Problem (mTSP) and TSP with neighbours (TSPN), we find trajectories of UAVs guaranteeing both reliable data collection and the shortest total flight times. The results include guidelines for WSN data collection design: K-means clustering provides minimum numbers of CHs to deploy. The genetic algorithm (GA) heuristic is near-optimal in determining trajectories of UAVs, with less than $3.5\%$ degradation. Hovering within a range of each CH improves a UAV's flight time. The environment type and the UAV's altitude impact data collection time. Finally, integrating trajectory fairness in multi-UAV large WSNs improves the data collection performance.   

The rest of the paper is organized as follows. Sections II and III present the system model and problem formulation respectively. In Section IV, the proposed solution is detailed. Section V illustrates the simulation results. Finally, Section VI concludes the study.

\begin{figure}[t]
	\centering
\includegraphics[width=210pt]{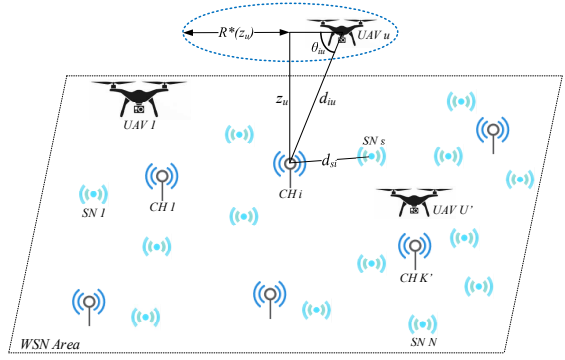}
	\caption{System model.}
	\label{Fig1}
\end{figure}

\section{System Model}
We assume a WSN network, where $N$ SNs are randomly located. SNs sense different kinds of data that they need to transmit to the network. We assume that several UAVs are deployed to collect SNs' data, as depicted in Fig. \ref{Fig1}. 
SNs are typically low power devices that cannot communicate directly with UAVs, hence $K$ CHs are deployed to collect data and send it to UAVs. We assume that a SN transmits its data to its associated CH using power $P_s$, while a CH communicates with a UAV with power $P_c> P_s$. Moreover, all communications are assumed to be orthogonal, i.e., no interference is occurring. 
Collection of data occurs as follows. After association with CH $i$, SN $s$ transmits its data. The received signal-to-noise ratio (SNR) at CH $i$, $\gamma_{i}^{ch}$, can be written as \vspace{-7pt} 

\small
\begin{equation}
    \label{eq:SNR_chi}
    \gamma_{i}^{\rm{ch}}=\frac{P_s d_{si}^{-\alpha}}{\sigma^2}, \forall i = 1, \ldots,K, \; \forall s =1,\ldots,N,
\end{equation}
\normalsize

\noindent
where $d_{si}$ is the distance between SN $s$ and CH $i$, $\alpha$ is the path-loss exponent, and $\sigma^2$ is the noise power. This communication link is considered successful if $\gamma_{i}^{\rm{ch}}\geq \gamma_{\rm{th}}$, where $\gamma_{th}$ is a chosen SNR threshold. Consequently, a maximum communication range for SN $s$ can be defined by \vspace{-7pt}

\small
\begin{equation}
    \label{eq:range_sn}
    d_{si} \leq d_{s}^{\rm{th}}= \left( \frac{P_s}{\sigma^2 \gamma_{\rm{th}}} \right)^{1/\alpha}, \forall s=1,\ldots,N.
\end{equation}
\normalsize

\noindent
Once data is received, CH $i$ waits for the UAV to hover above its position to send it $N_i$ data packets of size $D$ bits each. $N_i$ is equal to the number of sensors associated with CH $i$.  
We assume LoS and NLoS links, and no shadowing in the channel model. The latter is given by \cite{Hourani2014} \vspace{-6pt}

\small
\begin{equation}
\label{eq:avg_pathloss}
\Lambda_{iu}=\mathop{\mathbb{P}^{\rm{LoS}}_{iu}}\;l_{iu}^{\rm{LoS}}+\mathop{\mathbb{P}^{\rm{NLoS}}_{iu}}\;l_{iu}^{\rm{NLoS}},
\end{equation}
\normalsize

\noindent
where $\Lambda_{iu}$ is the average path-loss between CH $i$ and UAV $u$ ($i=1,\ldots,K$, $u=1,\ldots,U$), $\mathop{\mathbb{P}^{\rm{LoS}}_{i}}$ is the LoS probability, $\mathop{\mathbb{P}^{\rm{NLoS}}_{i}}=1-\mathop{\mathbb{P}^{\rm{LoS}}_{i}}$ is the NLoS probability, and $l_{iu}^{\rm{LoS}}$ and $l_{iu}^{\rm{NLoS}}$ are the LoS and NLoS path-losses. They are written as \vspace{-9pt}

\small
 \begin{equation}
 \label{eq:PLoS}
 \mathop{\mathbb{P}_{iu}^{\rm{LoS}}}=1/\left(1+ a \; e^{-b (\frac{180}{\pi}\theta_{iu}-a)}\right),
 \end{equation}
 \begin{equation}
 \label{eq:LLoS}
 l_{iu}^{m}=L_{\rm{FS}}(f_c)+20 \; \text{log}\left(d_{iu}\right)+\nu_{{m}}, \;\forall m \in \left\{ \text{\rm{LoS}}, \text{\rm{NLoS}} \right\}
 \end{equation}
 \normalsize

\noindent
where $a$ and $b$ are constants dictated by the environment, $d_{iu}$ and $\theta_{iu}$ are the distance (in $m$) and angle (in $rad$) between CH $i$ and UAV $u$ respectively, depicted in Fig. \ref{Fig1}. $L_{\rm{FS}}(f_c)=20 \;\text{log}\left(4\pi f_c /c\right)$, with $f_c$ is the carrier frequency and $c$ the light's velocity, and $\nu_{{m}}$ is the excessive path-loss coefficient. Using Friis formula, received power at UAV $u$ from CH $i$ is expressed by \vspace{-10pt}

\small
\begin{equation}
    \label{eq:Pr}
    P_{iu}^{\rm{UAV}}=P_c - \Lambda_{iu} \geq P_{\rm{th}}, \; \forall i=1,\ldots,K, \; \forall u=1,\ldots,U,
\end{equation}
\normalsize

\noindent
where $P_{\rm{th}}$ is the UAV's receiver sensitivity, i.e., above it, the communication between CH $i$ and UAV $u$ is successful.


In the next section, we formulate the joint problem of deploying a number of CHs and UAVs to collect data in the shortest time possible, while respecting budget (i.e., maximum number of deployable CHs and UAVs) and power constraints.

\section{Problem Formulation}
First, let $K' \leq K$ and $U' \leq U$ be the effective number of CHs and UAVs that are going to be deployed. Then, we define by $\mathcal{D}_i=\{\text{SN } s\;| \; d_{si}\leq d_s^{\rm{th}} \}$ the set of sensors associated with CH $i$, $\forall i=1,\ldots,K'$, and $\mathcal{C}_u=\{\text{CH } i \;| \; P_{iu}^{\rm{UAV}}\geq P_{\rm{th}} \}$ the set of CHs associated with UAV $u$, $\forall u=1,\ldots,U'$. The problem can be expressed as follows
\vspace{-15pt}

\small
\begin{subequations}
    \begin{align}
        \min_{\substack{\mathcal{D}_i,\mathcal{C}_u,\mathcal{W}_u,\mathcal{H}_u\\i=1,\ldots,K'\\ u=1,\ldots,U'}} & \quad 
	\frac{1}{U'}\sum_{u=1}^{U'} \sum_{i \in \mathcal{C}_u \cup \{0\}} \| \mathbf{w}_{u,i+1}-\mathbf{w}_{u,i} \|/V_u \tag{P1} \\
	\label{c1} 
	\text{s.t.}\quad & K' \leq K, \; U' \leq U,\tag{P1.a} \\
	\label{c2} & d_{si} \leq d_s^{\rm{th}}, \; \forall i=1,\ldots,K',\; s \in \mathcal{D}_i  \tag{P1.b}\\
	\label{c3}& {P_{iu}^{\rm{UAV}} \geq P_{\rm{th}}},\; \forall u=1,\ldots,U',\; \forall i \in \mathcal{C}_u, \tag{P1.c}\\
	\label{c4}& |\cup_{u=1}^{U'} \mathcal{C}_u|= K',\; \cap_{u=1}^{U'}\mathcal{C}_u= \emptyset, \tag{P1.d}\\
	\label{c5}& |\cup_{i=1}^{K'}\mathcal{D}_i|= N, \; \cap_{i=1}^{K'}\mathcal{D}_i= \emptyset, \tag{P1.e}
    \end{align}
\end{subequations}
\normalsize

\noindent
where $\mathcal{W}_u=\{\textbf{w}_{u}=[x_{u,i}, y_{u,i}, z_{u,i}]\;|\; \forall i \in \mathcal{C}_u\}$ is the set of ordered locations to visit by UAV $u$ (route) to collect data from its CHs, and $\textbf{w}_{u}$ is the location in the Cartesian coordinates system. Similarly,  $\mathcal{H}_u=\{\textbf{h}_{i}=[x_{i}, y_{i}, z_{i}]\;|\; \forall i \in \mathcal{C}_u\}$ is the set of ordered locations of CHs to visit by UAV $u$. Moreover, $\textbf{w}_{u,0}=\textbf{w}_{u,|\mathcal{C}_u|+1}$ is UAV $u$'s dockstation, and $V_u$ is UAV $u$'s average flight speed. Finally $|.|$ and $\|.\|$ are the Cardinality and Euclidean norm operators respectively.
The objective function minimizes the average data collection time, where we assume that communication time is fixed and long enough to collect all data from the CH successfully, hence ignored in this expression. Constraints (\ref{c1}) are the budget limitations in terms of number of CHs and UAVs. Whereas, (\ref{c2})-(\ref{c3}) and (\ref{c4})-(\ref{c5}) guarantee successful communications and associations between the CHs--SNs and UAVs--CHs, respectively.


%

The formulated problem is NP-hard. Indeed, in the special case of one UAV and already deployed CHs with respect to (\ref{c4}), the problem is reduced to finding the shortest UAV route. The latter can be comprehended as the TSP. In TSP, a salesman needs to visit a number of cities, while minimizing the traveled distance. Logically, the salesman and cities are assimilated by UAV and CHs respectively. Since TSP is NP-hard, then by restriction, our problem is also NP-hard.   


\section{Proposed Solution}
Solving problem (P1) directly is very difficult. Hence, we opt for a two-step approach as follows: 1) For given $K'$ and $U'$, we find the locations of CHs to satisfy (\ref{c2}), (\ref{c5}), i.e., sets $\mathcal{D}_i$ ($i=1,\ldots,K'$) and $\mathcal{H}=\cup_{u=1 }^{U'}\mathcal{H}_u$. This step is the WSN nodes clustering. 2) Then, 
for each UAV, the associated CHs, the trajectory and exact aerial locations to visit are determined, i.e., $\mathcal{C}_u$, $\mathcal{H}_u$ and $\mathcal{W}_u$, $\forall u=1,\ldots,U'$. This step is the trajectory planning. It is to be noted that the proposed approach yields a sub-optimal solution. Nevertheless, it  presents interesting implementation characteristics such as simplicity and full control of parameters.

\subsection{WSN Nodes Clustering}
Clustering SNs and deploying CHs to collect data is a widely used technique, as it is considered energy-efficient. Indeed, low-power SNs either cannot communicate directly with the collector (ex: UAV) or the latter has to get very close to SNs to collect data, which is energy wasting. Also, using CHs permits the centralization and filtering of data before transmitting it to the network. By doing so, CHs are able to process huge amounts of data and guarantee the scalability of the WSN. Moreover, clustering tolerates failures and malfunctions through filtering, backup CHs, and re-clustering. Finally, by adequately deploying CHs, data loads can be balanced among CHs, in order to extend their availability in the WSN \cite{Afsar2014}.   
In the literature, several approaches exist, such as K-means \cite{Kanungo2002}, Mean-Shift, Density-Based Spatial clustering, etc. In this paper, we opt for K-means to cluster SNs and place dedicated CHs. Hence, the associated clustering problem is 
\vspace{-13pt}

\small
\begin{subequations}
	\begin{align}
	\min_{\mathcal{H}, \mathcal{D}_i} & \quad 
	 \sum_{i=1}^{K'} \sum_{s \in \mathcal{D}_i} \| \textbf{g}_s - \textbf{h}_i \|^2   \tag{P2} \\
	\label{c21} 
	\text{s.t.}\quad & \text{(\ref{c1})},\;\text{(\ref{c2})}, \; \text{(\ref{c5})} \nonumber
	\end{align}
\end{subequations}
\normalsize
where $\mathbf{g}_{s}=[x_s,y_s,z_s]$ is the location of SN $s$, $\forall s=1,\ldots,N$. This problem is known to be NP-hard. K-means solves (P2) iteratively by updating the locations of CHs, as the averaged locations among SNs in the same cluster. 
The K-means method is inspired from \cite{Kanungo2002}. However, to be adapted to our system, i.e. to satisfy additional constraints (\ref{c2}) and (\ref{c5}), we integrate it into proposed Algorithm \ref{Algo1}.


\begin{algorithm}[h]
\small{
\caption{Determining the number and locations of CHs}
\label{Algo1}
\begin{algorithmic}[1]
\State Initialize $K'=1$
\State Initialize $d_s^{\rm{th}}$ using (\ref{eq:range_sn})
\State Run K-means \cite{Kanungo2002}, get $\mathcal{H}$ and $\mathcal{D}_1$
\State Calculate $d_{\rm{max}}=\max_{s \in \mathcal{C}_1} \| \mathbf{g}_s - \mathbf{h}_1 \|$
\While{$d_{\rm{max}}> d_s^{\rm{th}}$ and $K'< K$} 
    \State $K'=K'+1$
    \State Run K-means \cite{Kanungo2002}, get $\mathcal{H}$ and $\mathcal{D}_i$ ($i=1,\ldots,K'$)
    \State Calculate $d_{\rm{max}}=\max_{\substack{s \in \mathcal{D}_i\\ i=1,\ldots,K'}} \| \mathbf{g}_s - \mathbf{h}_i \|$
  \EndWhile
\State Return $\mathcal{H}$ and $\mathcal{D}_i$ ($i=1,\ldots,K'$)
\end{algorithmic}}
\end{algorithm}
\normalsize
 
\subsection{Trajectory Planning}
With the minimum number of CHs deployed, (P1) is reduced to associating CHs to UAVs and finding best trajectories to collect data. Each UAV travels from its dockstation to associated CHs, hovers to collect data, and returns to its dockstation once all CHs have been visited. This problem is known as the mTSP. Despite its NP-hardness, several algorithms have been proposed in the literature \cite{laporte1992}. 
First, we investigate trajectory optimization when UAVs hover exactly above CHs. Then, the case where UAVs hover within a range of CHs will be studied. The latter is known as TSPN \cite{Arkin1994}. 

\subsubsection{UAVs hover exactly above CHs}
For simplicity's sake, we assume that a UAV hovers at the same altitude for the whole trajectory, and that it hovers exactly above its associated CHs, such that constraint (\ref{c4}) is respected. Moreover, we assume that hovering time is fixed and long enough to collect all data from the CH successfully, 
and that UAVs have the same average flying speed $V_u=V$, $\forall u=1,\ldots,U'$. Hence, the trajectory planning problem can be written as \cite{BEKTAS2006} \vspace{-15pt}

\small
\begin{subequations}
	\begin{align}
	\min_{\mathcal{X}} & \quad 
	 \sum_{u=1}^{U'} \sum_{i =0}^{K'} \sum_{\substack{j=0\\j \neq i}}^{K'} \| \textbf{h}_i - \textbf{h}_j \| x_{ij}^u  \tag{P3} \\
	\label{c31} 
	\text{s.t.}\quad & \sum_{u=1}^{U'} \sum_{i=0}^{K'} x_{ij}^u =1,\; \forall j=0,\ldots,K', \; j\neq i \tag{P3.a} \\
	\label{c31_1} & \sum_{i=1}^{K'} x_{ip}^u-\sum_{j=1}^{K'}x_{pj}^u=0,\; \forall p=1,\ldots,K' \tag{P3.b},\; \forall u\\
	\label{c31_2} & \sum_{j=1}^{K'} x_{0j}^u=1, \;\forall u=1,\ldots,U' \tag{P3.c}\\
	\label{c31_3} & n_i - n_j + K' \sum_{u=1}^{U'} x_{ij}^u \leq K' -1 \tag{P3.d}\\
	\label{c32}& x_{ij}^u \in \left\{ 0,1\right\}, \tag{P3.d} 
	\end{align}
\end{subequations}
\normalsize
where $\| \textbf{h}_i - \textbf{h}_j \|$ is the distance between CHs $i$ and $j$, $\mathbf{h}_0=\textbf{w}_0=\textbf{w}_{u,0}$ is the initial and same dockstation for all UAVs, $\mathcal{X}=\{x_{ij}^u\;|\; i=1,\ldots,K',\; j=1,\ldots,K',\; u=1,\ldots,U' \}$ with $x_{ij}^u$ is a binary indicator of UAV $u$ traveling from CH $i$ to CH $j$, $x_{ij}^u=1$ if path is included in UAV $u$'s trajectory, otherwise, $x_{ij}^u=0$. Also, $n_i$ and $n_j$ are non-negative integers \cite{BEKTAS2006}. (\ref{c31}) state that each CH is visited exactly once. (\ref{c31_1}) are the flow conservation constraints, i.e., when a UAV visits a CH, then it must depart from the same CH. (\ref{c31_2}) ensures that a UAV is used exactly once, and (\ref{c31_3}) are the MTZ-based subtour elimination constraints, where degenerate tours not connected to the dockstation are omitted.

Since we assume that the visited locations of the UAVs are directly above the CHs, then a decided UAV location along the trajectory can be written as $\textbf{w}_u=[x_i,y_i,z_u]$, where the UAV is above CH $i$, and $z_u$ is the UAV's flying altitude. Using the obtained solution $\mathcal{X}$, $\mathcal{C}_u$ and $\mathcal{W}_u$ can then be deduced. 

Brute-force search explores all combinations of routes and is the most direct solution. However, it is computationally expensive and its complexity is high. The latter is given by
\small
\begin{equation}
    \label{eq:complex_BF}
    \mathcal{O}\left( \sum_{k_1=0}^{K'} \mathbb{C}_{K'}^{k_1}\; k_1! \sum_{k_2=0}^{K'}\ldots \sum_{k_{U'}=0}^{K'} \left( \prod_{i=2}^{U'} \mathbb{C}_{K'- \sum \limits_{j=1}^{U'-1} k_j}^{k_i}\; k_i! \right) \right),
\end{equation}
\normalsize
where $\mathbb{C}_i^j$ is the combination operation. 
Also, other proposed exact algorithms are time consuming and impractical for small computers and micro-controllers \cite{laporte1992}. Nonetheless, approximate and heuristic algorithms are interesting alternatives, especially if resulting routes are near-optimal. Consequently, we adopt in this paper two approximate trajectory planning approaches, namely the nearest neighbor (NN) and genetic algorithm (GA).
\begin{itemize}
    \item Using NN algorithm, a UAV starts by selecting the closest CH as its first destination. In the next steps, the closest CH not yet visited will be selected until all CHs are visited by the UAV. NN obtains routes quickly; however, their quality depends on initial locations . Its complexity for our problem is $\mathcal{O}\left( \left(K'+1\right)^2 \right)$.
    \item Genetic algorithms are introduced to solve combinatorial optimization problems. Several papers have used them to obtain the shortest paths in TSPs \cite{Larranaga1999}. 
    For our problem, the GA pseudo-algorithm is presented in Algorithm \ref{Algo2}. GA population is a set of trajectories for all UAVs, i.e., combinations of $\{\mathcal{C}_1,\ldots, \mathcal{C}_{U'}\}$ (and $\{ \mathcal{W}_1,\ldots, \mathcal{W}_{U'}\}$). A gene is a CH to be visited. An individual is the $U'$ trajectories satisfying (\ref{c31})-(\ref{c32}). The parents are combined solutions. The mating pool is a collection of parents to create the next generation. Fitness tells how short the sum of all trajectories is. Mutations introduce variations in the population by randomly swapping CHs in a given trajectory. Finally, elitism carries the best individuals to the next generation. GA's complexity can be given by $\mathcal{O}\left( \zeta \lambda  \right)$, where $\zeta$ is the population size and $\lambda$ the maximum number of generations. These parameters can be controlled with respect to the solution execution time trade-off, i.e., large $(\zeta,\lambda)$ gets near-optimal solutions at the expense of longer execution times, versus small $(\zeta,\lambda)$ for sub-optimal solutions, but in short execution times.
\end{itemize}

\begin{algorithm}[h]
\small{
\caption{GA Pseudo-Algorithm}
\label{Algo2}
\begin{algorithmic}[1]
\State Create population (i.e., combinations of $\{\mathcal{C}_1,\ldots, \mathcal{C}_{U'}\}$)
\State Determine fitness $\sum_{u=1}^{U'}\sum_{i \in \mathcal{C}_{u}\cup \{0\}} \| \mathbf{w}_{u,i+1}-\mathbf{w}_{u,i} \|$
\State Select mating pool (new combinations of $\{\mathcal{C}_1,\ldots, \mathcal{C}_{U'}\}$)
\State Create next generation using ordered crossover and elitism
\State Use swap mutation to introduce new combinations
\State Repeat steps 1-5 until convergence 
\State Return $\{\mathcal{C}_1,\ldots, \mathcal{C}_{U'}\}$ and associated $\{\mathcal{W}_1,\ldots, \mathcal{W}_{U'}\}$
\end{algorithmic}}
\end{algorithm}

\subsubsection{UAVs hover within a range of CHs}
A UAV can settle for a further location, but close enough to collect data. This can be seen as the CH having an aerial coverage area, in which communication to UAVs is successful. In order to determine the maximum radius of this coverage area for a given UAV altitude, we solve $P_{iu}^{\rm{UAV}}=P_c - \Lambda_{iu}$.   

\begin{Lemma}
\label{Lemma1}
Given the altitude $z$ of a UAV, the maximum coverage radius of a CH is expressed by \small
\begin{equation}
    \label{eq:radius}
    R^{*}(z)=\frac{z}{{\rm{tan}}\left[ f^{-1} \left[ P_c - P_{\rm{th}} - 20 \; \text{log}\left( z \right) - L_{\rm{FS}}\left( f_c \right)  \right] \right]},
\end{equation}
\normalsize
where $\rm{tan}(.)$ is the tangent function, and $f^{-1}$ is the inverse function of $f$ defined as
\small
\begin{equation}
    \label{eq:functF}
    f(\theta)=\frac{\nu_{\rm{LoS}}+ \nu_{\rm{NLoS}} \; a \; e^{-b\left( \theta-a \right)}}{1+a \; e^{-b\left( \theta-a \right)}}-20 \text{log}\left( \rm{sin}(\theta)\right),\; \theta \in [0,\pi].
\end{equation}
\normalsize
\end{Lemma}
\begin{IEEEproof}
For clarity, the designations $\Lambda_{iu}$, $d_{iu}$, $\theta_{iu}$ and $z_u$ are simplified into $\Lambda$, $d$, $\theta$ and $z$ respectively. Using (\ref{eq:avg_pathloss})-(\ref{eq:LLoS}), and $d=\frac{z}{\rm{sin}(\theta)}$, we obtain after some mathematical manipulations
\small
\begin{eqnarray}
\Lambda  &=& P_c - P_{\rm{th}} \Leftrightarrow \nonumber \\
20 \; \text{log}(z) - 20 \; \text{log}(\rm{sin}(\theta)) &+& \frac{\nu_{\rm{LoS}} +\nu_{\rm{NLoS}}\; a \; e^{-b (\theta - a)}}{1 + a \; e^{-b (\theta - a)}}  \nonumber \\ 
&+& L_{\rm{FS}}(f_c)= P_c - P_{\rm{th}} \nonumber \Leftrightarrow  \\
\label{eq:demo}
\frac{\nu_{\rm{LoS}} +\nu_{\rm{NLoS}}\; a \; e^{-b (\theta - a)}}{1 + a \; e^{-b (\theta - a)}} &-& 20 \; \text{log}(\rm{sin}(\theta)) =  P_c - P_{\rm{th}}\nonumber \\ 
    &-& L_{\rm{FS}}(f_c) - 20 \; \text{log}(z).
\end{eqnarray}
\normalsize
Let the function $f(\theta)$ be the left side of the equality (\ref{eq:demo}), hence the value of $\theta$ that achieves equality is given by
\small
\begin{equation}
\label{eq:theta}
    \theta = f^{-1}\left( P_c -P_{\rm{th}}- L_{\rm{FS}}(f_c) - 20 \text{log}(z) \right).
\end{equation}
\normalsize
It is to be noted that due to the complexity of $f(\theta)$, the value $\theta$ in (\ref{eq:theta}) needs to be determined numerically.
Since ${\rm{tan}}(\theta)=\frac{z}{R}$, where $R$ is the coverage radius, (\ref{eq:radius}) can be then obtained.
\end{IEEEproof}

\noindent
Now, given the maximum coverage radius of each CH, UAVs can optimize their trajectories by visiting the edges of CHs' coverage areas. This problem is identified as multiple-TSPN (mTSPN). To solve this problem, we propose a modification to the original mTSP solution. Indeed, after obtaining the trajectory among CHs to be visited as previously, we modify the hovering locations by adjusting them as the closest points on the coverage area edges. For instance, we assume that UAV $u$ is at coordinates $\textbf{w}_u=[x_u,y_u,z_u]$ and that the projection of the next CH to visit on the UAV's plane has coordinates $\bar{\textbf{w}}_c=[x_c,y_c,z_u]$. The associated CH's coverage perimeter can be expressed by the function \vspace{-5pt}

\small
\begin{equation}
    \label{eq:circle}
    \left( y-y_c \right)^2+\left( x-x_c \right)^2= (R^*(z_u))^2.
\end{equation}
\normalsize
Since the UAV flies in the direction of the CH to be visited, then the closest point on the edge of the coverage area is the first intersecting point between the line drawn through locations $\textbf{w}_u$ and $\bar{\textbf{w}}_c$ and the coverage perimeter. 
\begin{Lemma}
\label{Lemma2}
Let $y=p_1 \; x + p_2$ be the line's function, where $p_1=\frac{y_u - y_c}{x_u-x_c}$ and $p_2= y_u - p_1 \; x_u$. Then, the coordinates of the hovering location $\textbf{w}'_u$ can be given by 
\small
\begin{equation}
\label{eq:wup}
  \textbf{w}'_u=\begin{cases}
    \textbf{w}_u^0=[x_0,y_0,z_u],& \text{if } \|\textbf{w}_u^0 - \textbf{w}_u \| \leq \|\textbf{w}_u^1 - \textbf{w}_u \|  \\
    \textbf{w}_u^1=[x_1,y_1,z_u], & \text{otherwise}
\end{cases}
\end{equation}
\normalsize
\end{Lemma}
where $x_0=\frac{-q_2 + \sqrt{q_2^2-q_1\;q_3}}{q_1}$, $x_1=-\frac{q_2 + \sqrt{q_2^2-q_1\;q_3}}{q_1}$, $y_0=p_1\;x_0+p_2$, $y_1=p_1\;x_1+p_2$, $q_1={p_1}^2 +1$, $q_2=p_1 (p_2-y_c)-x_c$ and $q_3=x_c^2-{(R^*(z_u))}^2$.

\begin{IEEEproof}
The intersection points between the line drawn through $\textbf{w}_u$ and $\bar{\textbf{w}}_c$ and the coverage perimeter satisfy both (\ref{eq:circle}) and $y=p_1 \; x + p_2$. Hence, by substituting $y$ of the line's equation into (\ref{eq:circle}), we get after some manipulations \vspace{-10pt}

\small
\begin{eqnarray}
\label{eq:demo3}
&&(p_1  x + (p_2-y_c))^2 + (x-x_c)^2=(R^*(z_u))^2 \\
&\Leftrightarrow& x^2 \underbrace{(p_1^2 + 1)}_{=q_1} + 2 x \underbrace{(p_1 (p_2-y_c)-x_c)}_{=q_2} + \underbrace{x_c^2-(R^*(z_u))^2}_{= q_3}=0. \nonumber
\end{eqnarray}
\normalsize

\noindent
Thus, (\ref{eq:demo3}) is a polynomial that can be solved as in (\ref{eq:wup}).
\end{IEEEproof}

\section{Simulation Results}
We assume an area of $10 \times 10$ km$^2$, populated with $N=500$ SNs. We fix the following parameters $\gamma_{\rm{th}}=10^{-4}$, $f_c=2$ GHz, $c=3.10^8$ m/s, $P_c=20$ dBm and $P_{\rm{th}}=-100$ dBm. Moreover, we assume that $d_s^{\rm{th}} \in [100, 2900]$ meters and $\alpha \in [2,5]$, hence $\frac{P_s}{\sigma^2}=\left(d_s^{\rm{th}}\right)^\alpha \cdot \gamma_{\rm{th}}$. Also, we consider for GA $\zeta=500$ and $\lambda=1000$. Unless otherwise stated, the flying altitude of UAVs is 200 m and the environment parameters are for the urban one $(a,b)=(9.6117, 0.739)$ \cite{Hourani2014}.

\begin{figure*}[t]
     \begin{minipage}{0.315\linewidth}
     \includegraphics[width=175pt]{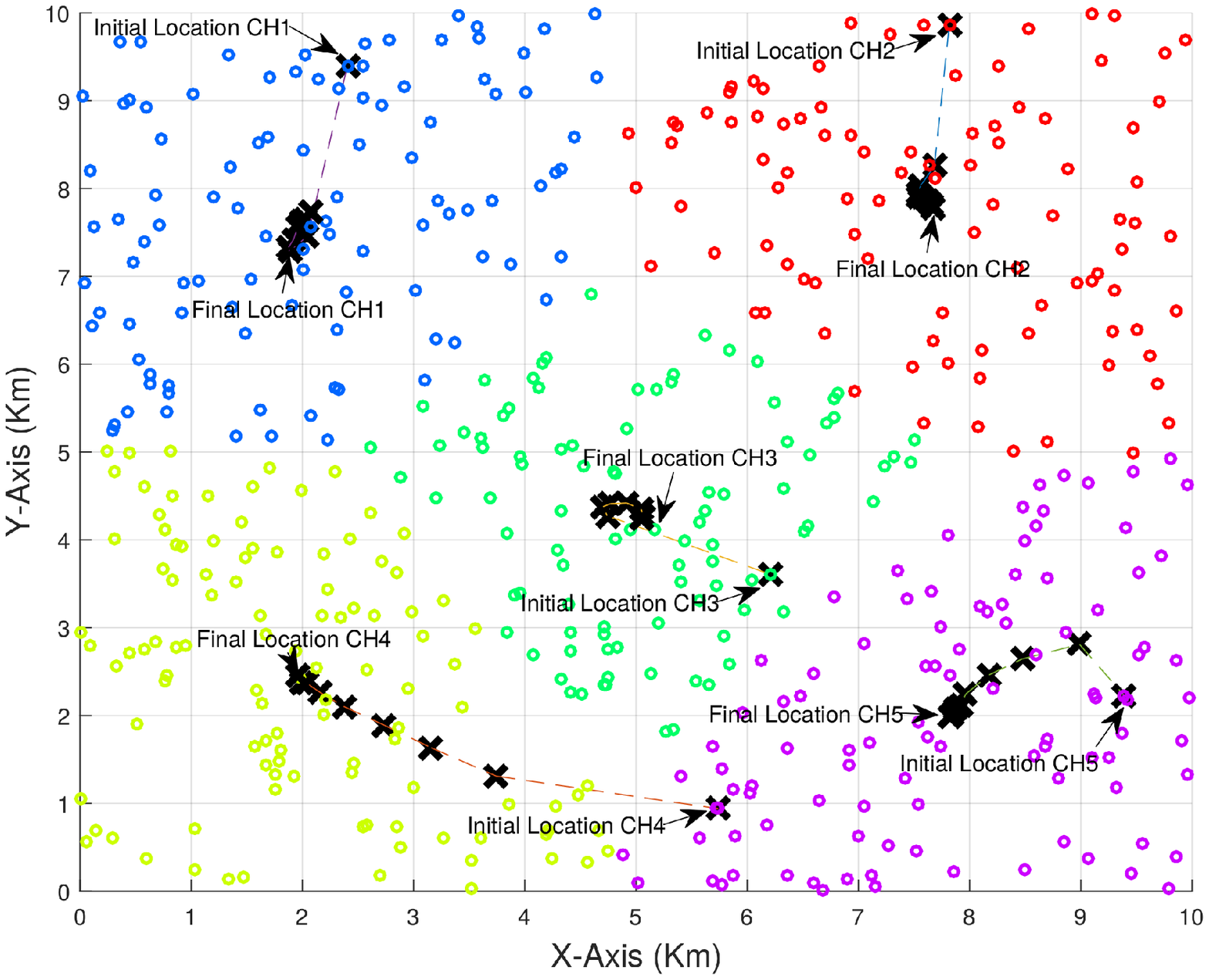}
       \caption{Convergence of K-means (10 iterations).}
    \label{Fig1:Kmeans}
     \end{minipage}
     \hfill
     \begin{minipage}{0.315\linewidth}
   \includegraphics[width=175pt]{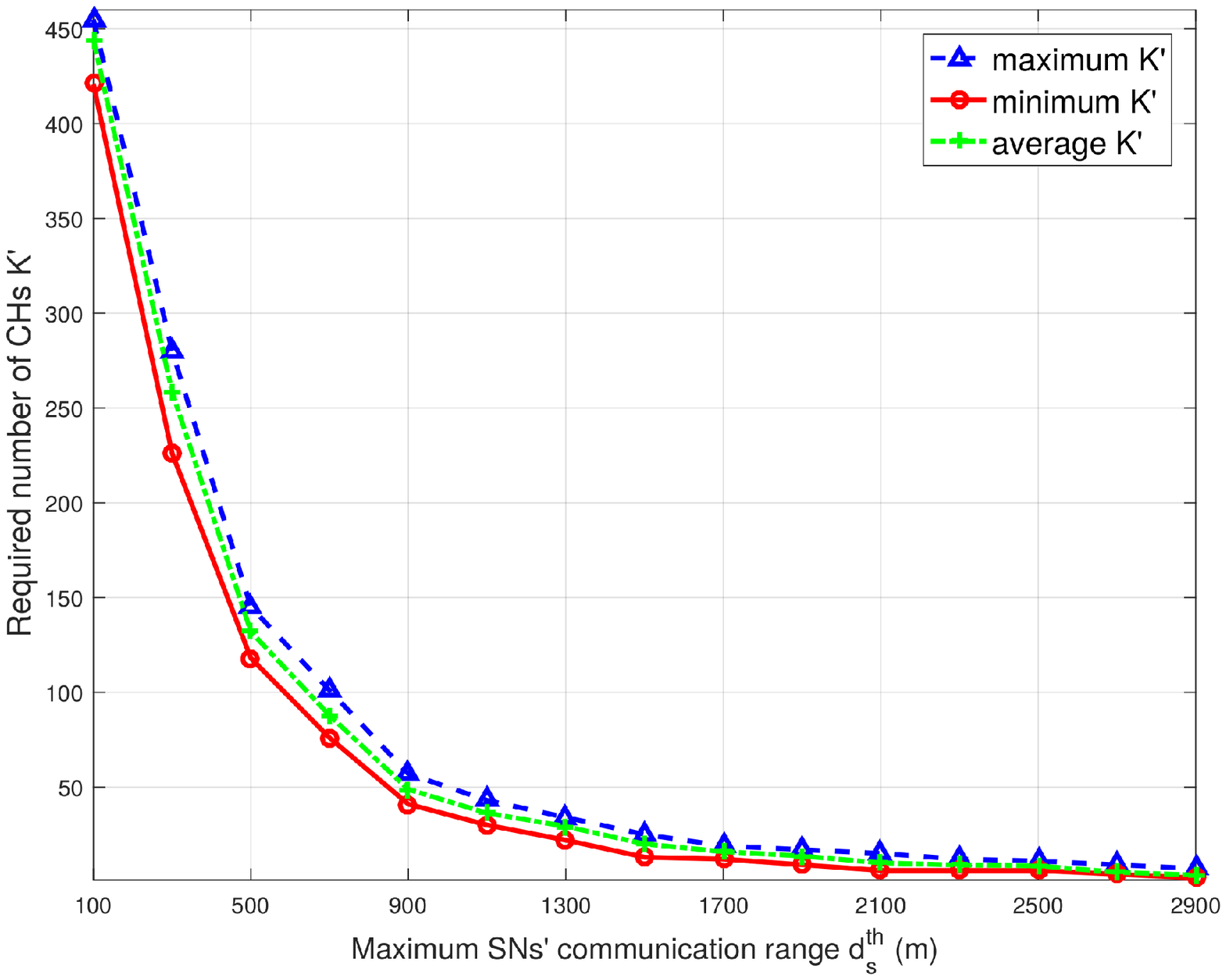}
	\caption{Required number of CHs vs. communication range of SNs.}
	\label{Fig2}
     \end{minipage}
     \hfill
     \begin{minipage}{0.315\linewidth}
    \includegraphics[width=160pt]{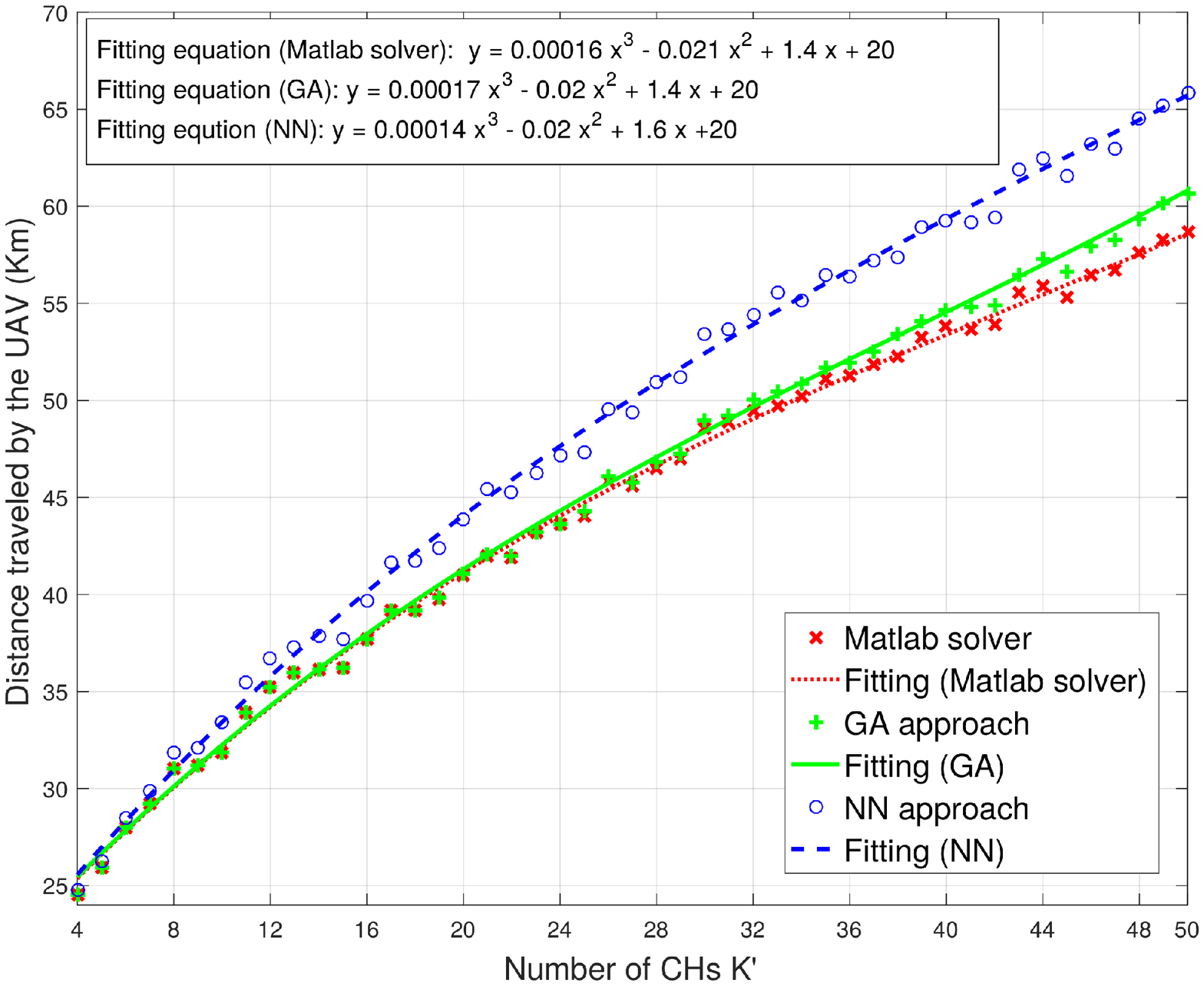}
	\caption{Distance traveled by the UAV vs. number of CHs.}
	\label{Fig3}
	\end{minipage}
   \end{figure*}



In Fig. \ref{Fig1:Kmeans}, we show how K-means clustering is processed to find the locations of CHs, such as (\ref{c2}) is satisfied. Given $K'=5$, the locations of CHs are updated in every iteration until the final locations are obtained. Identically colored SNs (circles) are associated with the CH in their region.  


Due to initialization randomness, K-means converges often to sub-optimal solutions. In order to evaluate its influence on our system, we run Algorithm \ref{Algo1} for several times (x100) and for different communication ranges  $d_s^{\rm{th}} \in [100, 2900]$. The results are plotted in Fig. \ref{Fig2}. As shown, with $d_s^{\rm{th}}$ increasing, a smaller number of CHs is needed. However, for $d_s^{\rm{th}}< 1000$ m, it is critical to have a very large number of CHs in order to guarantee communications to all SNs. Also, for a given $d_s^{\rm{th}}$, $K'$ can be different depending on the obtained locations of CHs. For instance, $K'$ has to be between 22-34 CHs for $d_s^{\rm{th}}=1700$m, to associate all SNs with CHs. 

Given $U'=1$, we compare in Fig. \ref{Fig3} the performances of the optimal solution (obtained using Matlab solver), proposed GA algorithm, and NN algorithm, in terms of the UAV's traveled distance, versus the number of CHs to visit. We assume here that the UAV hovers exactly above each CH. As $K'$ grows, distance traveled increases for all approaches. This is expected since visiting more CHs inevitably implies traveling longer distances. Moreover, Matlab solver achieves the shortest traveled distance, while GA follows closely with only 0\% to 3.5\% degradation. NN approach presents the worst performance with a degradation with $K'$ up to 12\%.   
Since GA achieves near-optimal performances, it will be used in the following simulations. 

\begin{figure*}[t]
     \begin{minipage}{0.315\linewidth}
       \includegraphics[width=175pt]{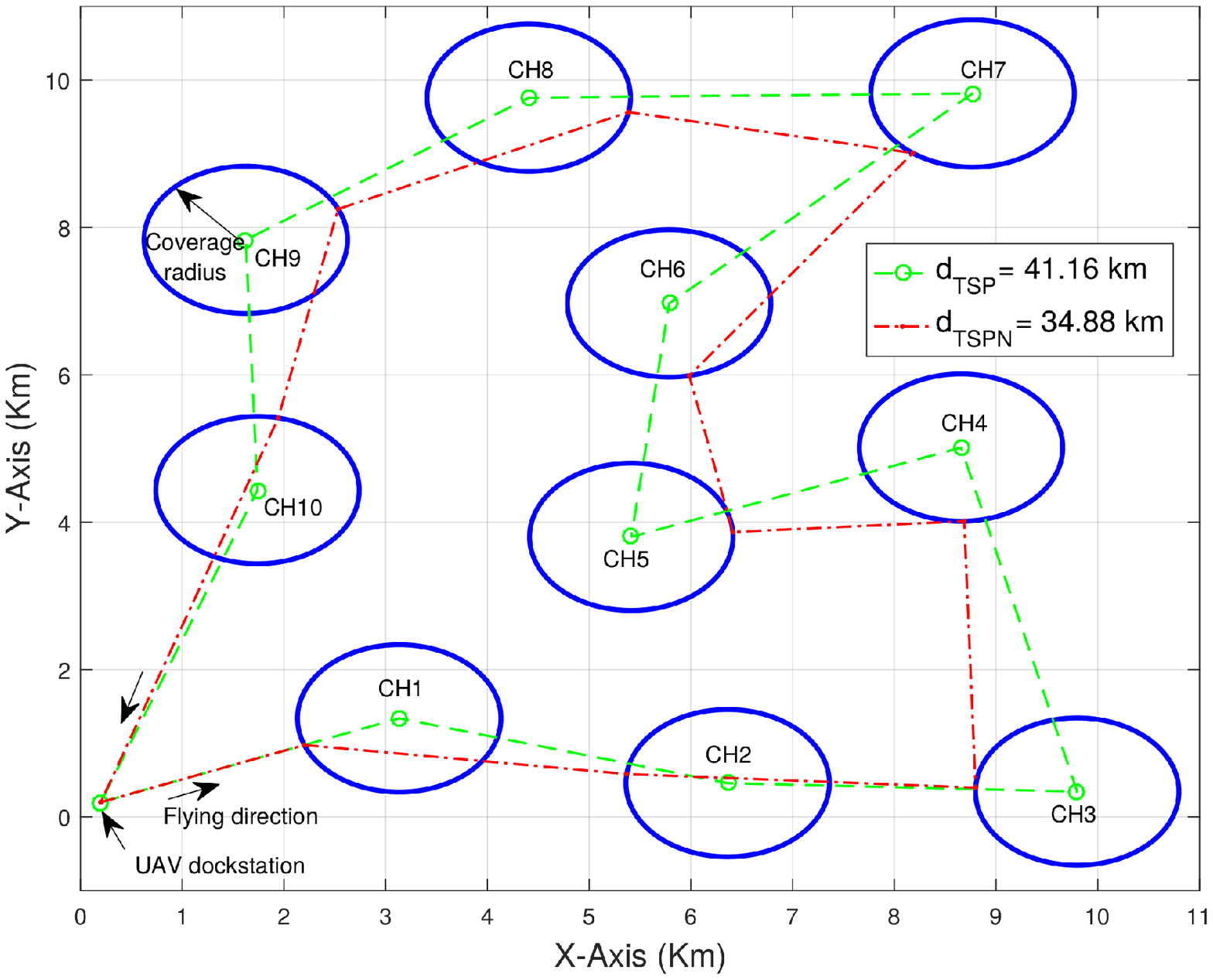}
       \caption{Trajectories of UAVs.}
        \label{Fig4}
     \end{minipage}
     \hfill
     \begin{minipage}{0.315\linewidth}
    \includegraphics[width=175pt]{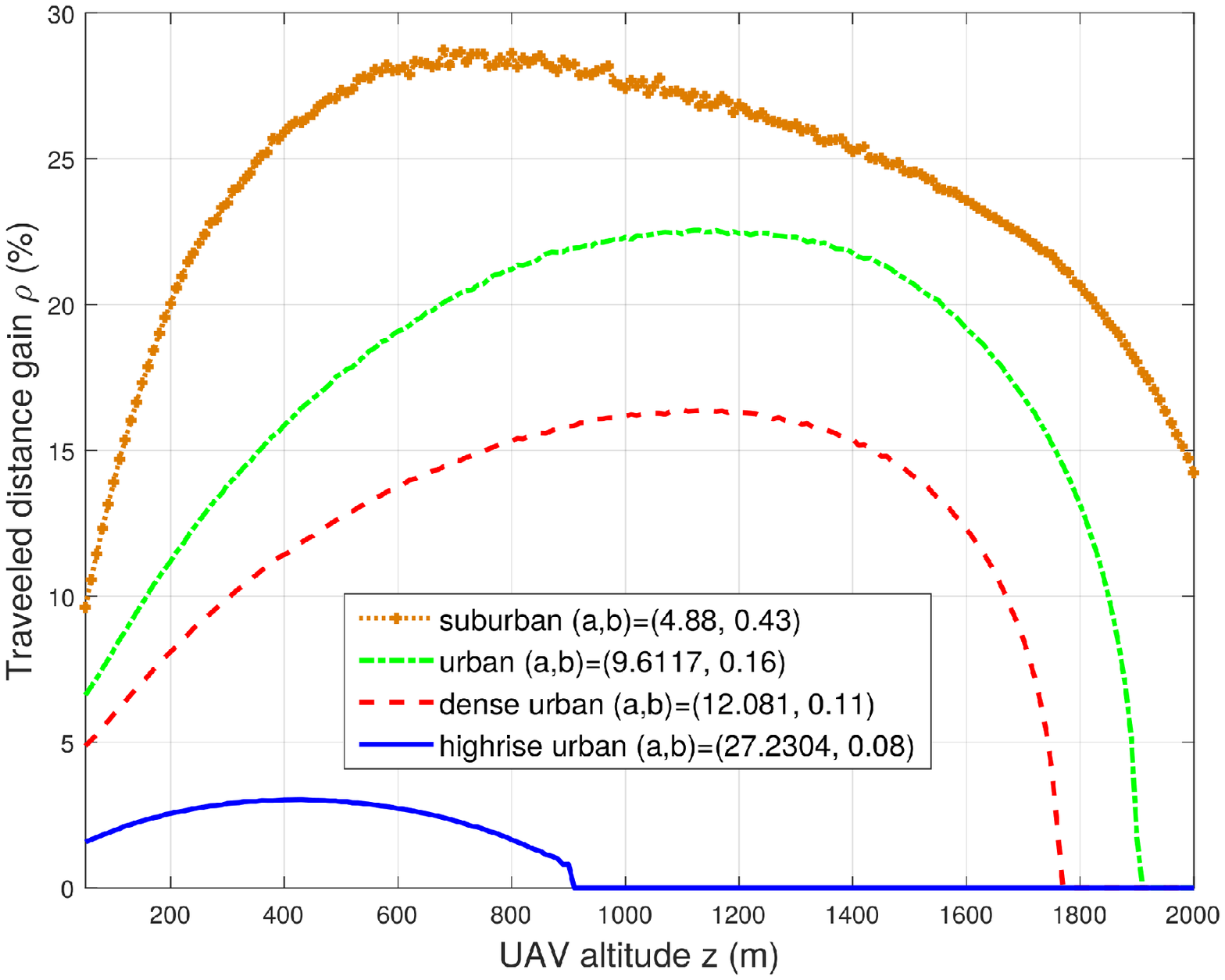}
	\caption{Average traveled distance gain vs. UAV altitude.}
	\label{Fig4_2}
     \end{minipage}
     \hfill
     \begin{minipage}{0.315\linewidth}
     \includegraphics[width=175pt]{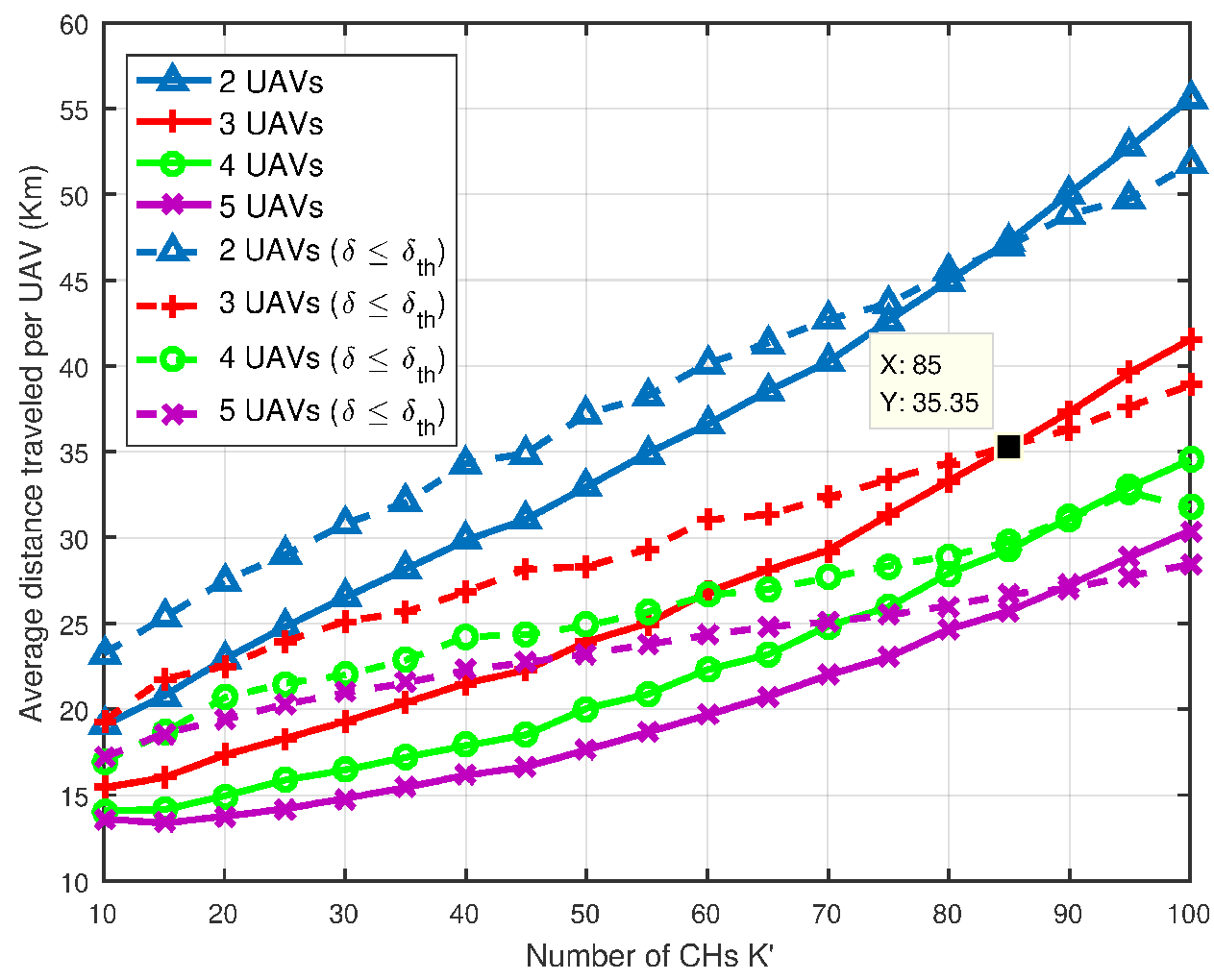}
	\caption{Average distance traveled per UAV vs. number of CHs.}
	\label{Fig5}
	\end{minipage}
   \end{figure*}

In Fig. \ref{Fig4}, we compare the distances traveled by UAVs when the latter hover either exactly above (TSP), or within a range of 1 km from each CH (TSPN). Let $d_{\rm{TSP}}$ and $d_{\rm{TSPN}}$ be the associated UAV's trajectory lengths respectively. As illustrated, TSPN achieves a shorter distance trajectory, while guaranteeing data collection from all 10 CHs. The traveled distance gain, defined by $\rho=1-\frac{d_{\rm{TSPN}}}{d_{\rm{TSP}}}$, can be calculated as $\rho=15.27\%$ in this scenario. That means, hovering within a distance of each CH is advantageous in shortening the distance traveled, and thus the data collection mission time.

In Fig. \ref{Fig4_2}, we illustrate the relation between the UAV altitude $z$ and the traveled distance gain $\rho$ for different environment types. The scenario has 20 CHs, and the environment parameters $(a,b)$, as defined in (\ref{eq:PLoS}), are given within the figure. Moreover, (\ref{eq:radius}) and (\ref{eq:wup}) are used to calculate the maximum coverage radius and the TSPN trajectory, respectively. As $z$ increases, $\rho$ increases at first until reaching a maximum value. The latter corresponds to the optimal UAV altitude, where CH-UAV communications satisfy the required $P_{\rm{th}}$. Beyond it, $\rho$ degrades rapidly as $z$ increases.
Indeed, a higher altitude degrades the CH-UAV communication link. The intersection point between any curve and the X-Axis presents the maximum altitude at which the UAV needs to hover exactly above the CH to collect data ($\rho = 0$). Above it, transmissions would fail. 
Also, $\rho$ is more important as the environment becomes less urbanized. Indeed, with better LoS links, communications are tolerated on wider coverage areas. 

\begin{figure}[t]
	\centering
	\includegraphics[width=178pt]{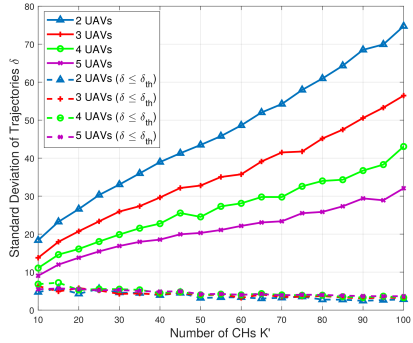}
	\caption{Standard deviation of trajectories vs. number of CHs.}
	\label{Fig6}
\end{figure}

In Figs. \ref{Fig5}-\ref{Fig6}, we investigate the performance of the multi-UAV system in terms of average distance traveled per UAV versus the number of CHs $K'$, and we compare the associated standard deviation of the trajectories, denoted $\delta$. As shown in Fig. \ref{Fig5}, the average distance traveled increases with $K'$. Moreover, as the number of UAVs $U'$ grows, the average distance traveled decreases significantly at first, then slowly. This means that adding a UAV to a system using initially a small number of UAVs is more beneficial than to a system using already a large number of UAVs. We notice in Fig. \ref{Fig6} that the standard deviation (solid lines) decreases when $U'$ increases, i.e., a system with a higher number of UAVs would achieve fairer trajectories among UAVs. 

Nevertheless, UAVs are usually constrained by their flight time (alternatively, their flight distance), consequently, an optimal but unfair (in terms of trajectory lengths) solution may not be adequate. In order to leverage fairness among multi-UAV trajectories, we propose to limit the standard deviation by a threshold $\delta_{\rm{th}}=10$. The associated results are given in Figs. \ref{Fig5}-\ref{Fig6} by the dashed lines. According to Fig. \ref{Fig5}, average traveled distance for $\delta \leq \delta_{\rm{th}}$ is higher than in the first case (solid lines), for $K'$ lower than a certain value (e.g., $K'=85$ for 3 UAVs). However, above it, the fair solution outperforms the first one. Indeed, since GA's parameters $(\zeta, \lambda)$ are fixed and $K'$ is increasing, Algorithm \ref{Algo2} can only provide sub-optimal solutions. Meanwhile, the condition $\delta \leq \delta_{\rm{th}}$ allows exploring other solutions that turn out to be more efficient. In Fig. \ref{Fig6}, we validate $\delta \leq \delta_{\rm{th}}$ in all scenarios (dashed lines).

\section{Conclusion}
In this paper, we proposed a framework for UAV-based WSNs, which aimed to collect data in the shortest time frame and at the lowest cost, in terms of deployed cluster heads and UAVs. Specifically, we used clustering to optimize the number and locations of cluster heads. Then, we leveraged optimal and heuristic solutions to mTSP and TSPN to obtain optimized number and trajectories of UAVs. Simulation results provide guidelines for data collection design in WSNs: the selection of the number of CHs and their clustering has to be carefully processed. The heuristic GA used to deploy and plan trajectories of UAVs was shown to be near-optimal with only $3.5\%$ degradation. Data collection times can be minimized by hovering within a range of visited CHs. Also, the data collection time is influenced by the environment and flight altitude. Finally, the integration of trajectory fairness is beneficial in large WSNs.

\bibliographystyle{IEEEtran}
\bibliography{IEEEabrv,tau}

\begin{thebibliography}{10}
\providecommand{\url}[1]{#1}
\csname url@samestyle\endcsname
\providecommand{\newblock}{\relax}
\providecommand{\bibinfo}[2]{#2}
\providecommand{\BIBentrySTDinterwordspacing}{\spaceskip=0pt\relax}
\providecommand{\BIBentryALTinterwordstretchfactor}{4}
\providecommand{\BIBentryALTinterwordspacing}{\spaceskip=\fontdimen2\font plus
\BIBentryALTinterwordstretchfactor\fontdimen3\font minus
  \fontdimen4\font\relax}
\providecommand{\BIBforeignlanguage}[2]{{%
\expandafter\ifx\csname l@#1\endcsname\relax
\typeout{** WARNING: IEEEtran.bst: No hyphenation pattern has been}%
\typeout{** loaded for the language `#1'. Using the pattern for}%
\typeout{** the default language instead.}%
\else
\language=\csname l@#1\endcsname
\fi
#2}}
\providecommand{\BIBdecl}{\relax}
\BIBdecl

\bibitem{Irem2016}
R.~I. Bor-Yaliniz, A.~El-Keyi, and H.~Yanikomeroglu, ``{‘Efficient 3-D
  placement of an aerial base station in next generation cellular networks},''
  in \emph{Proc. IEEE Int. Conf. Commun. (ICC)}, May 2016, pp. 1--5.

\bibitem{Alzenad2017}
M.~Alzenad, A.~El-Keyi, F.~Lagum, and H.~Yanikomeroglu, ``{3-D placement of an
  unmanned aerial vehicle base station (UAV-BS) for energy-efficient maximal
  coverage},'' \emph{IEEE Wireless Commun. Lett.}, vol.~64, no.~6, pp.
  434--437, Aug. 2016.

\bibitem{Pang2014}
Y.~{Pang}, Y.~{Zhang}, Y.~{Gu}, M.~{Pan}, Z.~{Han}, and P.~{Li}, ``{Efficient
  data collection for wireless rechargeable sensor clusters in harsh terrains
  using UAVs},'' in \emph{Proc. IEEE Global Commun. Conf.}, Dec. 2014, pp.
  234--239.

\bibitem{Mozaf2016}
M.~{Mozaffari}, W.~{Saad}, M.~{Bennis}, and M.~{Debbah}, ``{Unmanned aerial
  vehicle with underlaid device-to-device communications: performance and
  tradeoffs},'' \emph{IEEE Trans. Wireless Commun.}, vol.~15, no.~6, pp.
  3949--3963, Jun. 2016.

\bibitem{Zeng2018_1}
Y.~{Zeng}, X.~{Xu}, and R.~{Zhang}, ``{Trajectory design for completion time
  minimization in UAV-enabled multicasting},'' \emph{IEEE Trans. Wireless
  Commun.}, vol.~17, no.~4, pp. 2233--2246, Apr. 2018.

\bibitem{Zeng2019}
Y.~{Zeng}, J.~{Xu}, and R.~{Zhang}, ``{Energy minimization for wireless
  communication with rotary-wing UAV},'' \emph{IEEE Trans. Wireless Commun.},
  vol.~18, no.~4, pp. 2329--2345, Apr. 2019.

\bibitem{Mozaf2016_1}
M.~{Mozaffari}, W.~{Saad}, M.~{Bennis}, and M.~{Debbah}, ``{Mobile Internet of
  things: can UAVs provide an energy-efficient mobile architecture?}'' in
  \emph{Proc. IEEE Global Commun. Conf.}, Dec. 2016, pp. 1--6.

\bibitem{Afsar2014}
M.~Afsar and M.-H. Tayarani, ``{Clustering in sensor networks: a literature
  survey},'' \emph{Elsevier J. Netw. Comput. Appl.}, vol.~46, pp. 198--226,
  2014.

\bibitem{Hourani2014}
A.~{Al-Hourani}, S.~{Kandeepan}, and S.~{Lardner}, ``{Optimal LAP altitude for
  maximum coverage},'' \emph{IEEE Wireless Commun. Lett.}, vol.~3, no.~6, pp.
  569--572, Dec. 2014.

\bibitem{Kanungo2002}
T.~{Kanungo}, D.~M. {Mount}, N.~S. {Netanyahu}, C.~D. {Piatko}, R.~{Silverman},
  and A.~Y. {Wu}, ``{An efficient K-means clustering algorithm: analysis and
  implementation},'' \emph{IEEE Trans. Pattern Analysis and Machine
  Intelligence}, vol.~24, no.~7, pp. 881--892, Jul. 2002.

\bibitem{laporte1992}
G.~Laporte, ``{The vehicle routing problem: an overview of exact and
  approximate algorithms},'' \emph{European J. of Op. Research}, vol.~59,
  no.~3, pp. 345 -- 358, Jun. 1992.

\bibitem{Arkin1994}
E.~M. Arkin and R.~Hassin, ``{Approximation algorithms for the geometric
  covering salesman problem},'' \emph{Discrete Appl. Math.}, vol.~55, no.~3,
  pp. 197--218, Dec. 1994.

\bibitem{BEKTAS2006}
T.~Bektas, ``{The multiple traveling salesman problem: an overview of
  formulations and solution procedures},'' \emph{Omega}, vol.~34, no.~3, pp.
  209 -- 219, Jun. 2006.

\bibitem{Larranaga1999}
P.~{Larranaga}, C.~{Kuijpers}, C.~{Murga}, and al, ``{Genetic algorithms for
  the travelling salesman problem: a review of representations and
  operators},'' \emph{Spr. Artif. Intel. Rev.}, vol.~13, no.~2, pp. 129--170,
  Apr. 1999.

\end{thebibliography}

\end{document}